Yttrium-90 TOF-PET based EUD predicts response post liver radioembolizations using standard manufacturer reconstruction parameters.

Michel Hesse, Philipe d'Abadie, Renaud Lhommel, Francois Jamar, Stephan Walrand

Cliniques universitaires saint-Luc, Brussels, Belgium

Corresponding author: Stephan.walrad@uclouvain.be

**Abstract:**

*Purpose:* Explaining why 90Y TOF-PET based equivalent uniform dose (EUD) using standard FDG reconstruction parameters has been shown to predict response.

*Methods:* The hot rods insert of a Jaszczak deluxe phantom was partially filled with a 2.65 GBq $^{90}$Y - 300ml DTPA water solution resulting in a 100 Gy mean absorbed dose in the 6 sectors. A two bed 20min/position acquisition was performed on a 550ps- and on a 320ps- TOF-PET/CT and reconstructed with standard FDG reconstruction parameters, without and with additional filtering. The whole procedure was repeated on both PET after adding a 300ml of water (50Gy setup). The phantom was acquired again after decay by a factor 10 (5Gy setup), but with 200min per bed position. For comparison the phantom was also acquired with $^{18}$F activity corresponding to a clinical FDG whole body acquisition.

*Results:* The 100Gy-setup provided hot rod sectors image almost as good as in $^{18}$F phantom. However, despite acquisition time compensation the 5Gy-setup provides much lower quality imaging. TOF-PET based sectors EUDs for the three large rod sectors were in agreement with the actual EUDs computed with a radiosensitivity $0.021Gy^{-1}$ well in the range observed in external beam radiotherapy (EBRT), i.e. $0.01$-$0.04Gy^{-1}$. This agreement explains the reunification of the dose-response relationships of the glass and resin spheres in HCC using the TOF-PET based EUD. Additional filtering reduced the EUDs agreement quality.

*Conclusions:* Standard FDG reconstruction parameters are suitable in TOF-PET post $^{90}$Y liver radioembolization for accurate tumor EUD computation. The present results clearly rules out the use of low specific activity phantom studies to optimize reconstruction parameters.

## Introduction

The first dose-response dependence in liver radioembolization was obtained using [90]Y loaded resin spheres as early as 1994 by Lau et al. in a heroic way (1): normal liver and tumour doses were measured in the catheterization room by scanning the liver surface with a calibrated intraoperative beta probe. More remarkably, small sphere amounts were sequentially injected until the planned liver dose was reached according to the beta probe measure. The 30-months patient follow-up showed a clear splitting of the survival rate for a 120Gy tumours dose threshold.

Modern imaging based dose-response correlations in [90]Y liver radioembolization have initially been reported using tumour absorbed doses assessed with pre-therapy [99m]Tc-MAA SPECT (2). Later, more convincing relations between tumour control probability (TCP) and absorbed dose were obtained with post-radioembolisation [90]Y bremsstrahlung SPECT for resin spheres (3) as well as for glass spheres (4). Up to now, more dose-response correlations were reported with this last modality (5-8) than with [99m]Tc-MAA SPECT (9,10).

The first [90]Y PET/CT imaging in human in 2009 (11), triggered many phantom studies in order to assess the optimal PET reconstruction parameters in dose assessment post liver radioembolization (12-23). Contrast recovery and noise level were evaluated for various spheres diameters, or vials sizes, filled with homogeneous activities modelling tumours and surrounded by a homogenous active background modelling the healthy liver parenchyma. The specific activity ratio between modelled tumours and the background ranged from 4 to 8. The modelled liver specific activity was about 1/2 to 1/8 fold that reached in typical liver radioembolization. However, due to the constant randoms rate generated by the natural radioactivity of lutetium based crystals, a lower specific activity in the target of interest cannot fully be counterbalanced by increasing the acquisition time or by summing several slices (24-27).

Reported optimal reconstruction tradeoffs between contrast recovery and noise control ranged from 1 up to 3 iterations x 21 subset, with or without 5 mm FWHM Gaussian post filtering (12-23). This large variation in optimal reconstruction parameters results from the different chosen contrast recovery and noise control tradeoffs, and especially from the different phantom setups and from the presence or absence of TOF assessment.

Besides these phantom studies, the first reported [90]Y TOF-PET based dose-response correlation showed that, similarly to external beam radiotherapy (EBRT), the baseline haemoglobin level had a major impact on the absorbed dose efficacy (28). This impact was later confirmed in a retrospective analysis of 606 liver radioembolizations (29). Other [90]Y PET based dose relations were reported (30-

33) confirming the factor 2 ratio for efficacy and toxicity per Gy already observed in the bremsstrahlung SPECT studies between glass and resin spheres.

Recently, [90]Y TOF-PET based equivalent uniform dose (EUD) was shown to reunifie survival response observed for glass and resin spheres radioembolizations in HCC using the same 40Gy-dose threshold, similar to that used in EBRT (34). Besides giving a better understanding of the radiobiology underlining the tumour response in radioembolization, the EUD formalism takes automatically into account the dose heterogeneity inside the tumour. Indeed, tumour doses can exhibit very different heterogeneity levels depending on the tumour vascularisation, and also on the sphere specific activity that can be finely tuned by letting the device vial decay before the catheterization.

This reunification was obtained using the standard FDG reconstruction parameters, i.e. 3 iterations x 33 subsets without filtering, followed by a spatial resolution recovery. The purpose of this study is to evaluate on a phantom whether this reconstruction setup was really adapted for tumour EUD assessment in clinical statistics. In order to assess the reconstruction robustness versus the heterogeneity pattern, the EUD adequacy was evaluated in the six hot rod sectors of a Jaszczak deluxe phantom. The sectors insert was only partially filled in order to get a regional count rate similar to that of radioembolized tumours.

**Material and methods**

*Phantom setup*

A Jaszczak deluxe phantom containing the hot rod insert (cylinder diameters: 4.8, 6.4, 7.9, 9.5, 11.1, 12.7 mm) was imaged set in the vertical position (Fig. 1). The insert lay on 2mm-spacers set at the bottom of the phantom tank allowing a free communication of liquid between rods. The fixation holes of the insert were filled with solid perpex cylinders in order to avoid extra active rods in between the six sectors.

A 2.65 GBq [90]Y - 300ml DTPA water solution was poured in the phantom, giving a 18mm-height filling of the rods. This pouring was slowly performed using a 50 ml syringe connected to 2 catheter lines with a three way tap, the tip of one being in the [90]Y-DTPA container and the other one located in the 2mm-space below the sectors insert. This method allows avoiding any air bubble formation in the rods. The inner rods absorbed dose was 50x3/0.3 = 433Gy, resulting in a ≈100 Gy mean absorbed dose in each whole sector region, i.e. including rods and plastic. Indeed, in a periodic hexagonal mesh the rods volume fraction is 0.227 independently of the rod diameters.

The phantom was set on a 13x22x29 cm³ paper block of 0.91g/cm³ density. Taking into account the phantom cover and hot sectors thickness above the ⁹⁰Y solution and the phantom bottom wall, this results in an equivalent 21cm length water attenuation in the vertical direction similar to that of the fully filled phantom set in the conventional horizontal position.

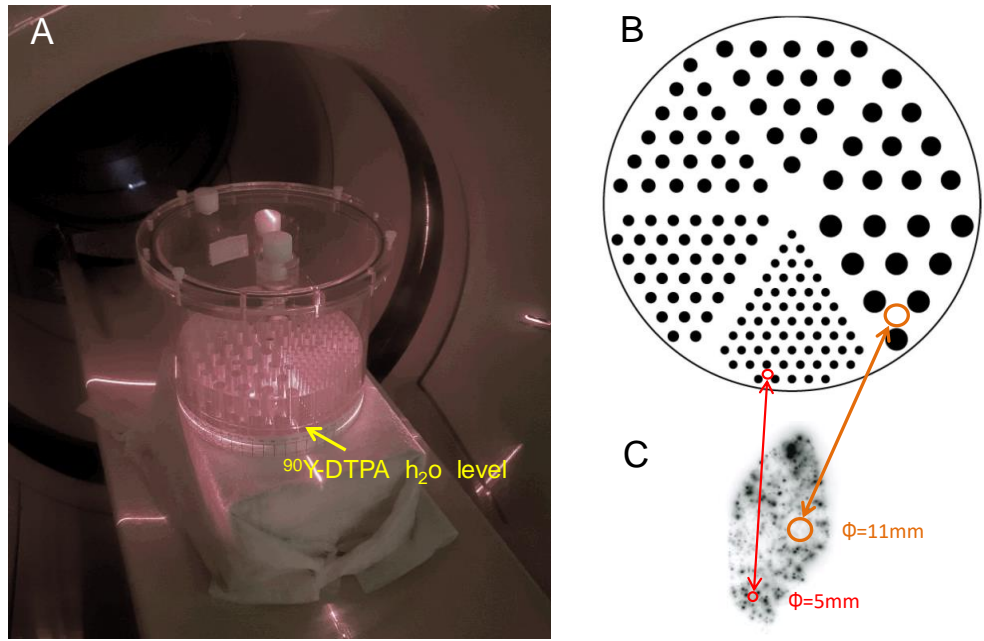

FIGURE 1: A) Jaszczak deluxe phantom set in vertical position on a paper bloc modelling the patient attenuation. Only a part of the hot rod insert was filled with active solution in order to reach a typical clinical tumour absorbed dose using a 3GBq total ⁹⁰Y activity. B) Actual hot rods map. C) Autoradiography of a normal liver (NL) tissue resected 9 days post ⁹⁰Y resin spheres liver radioembolization delivering 52Gy to the NL tissue (reprinted from (35) with permission of EJNMMI). B and C are represented at the same scale.

*Acquisitions*

A two bed 20min/position acquisition was performed on a 550ps TOF-PET/CT (Philips Gemini TF64 (36)) and on a 320ps TOF-PET/CT (Philips Vereos (37)). Standard FDG reconstruction parameters advised by the manufacturer were used, i.e. 4x4x4mm3 reconstruction voxel with 3 iterations x 33 subsets for the 550ps TFO-PET and 3 iterations x 15 subsets for the 320ps TOF-PET. For comparison purpose, EUDs were also computed after applying a 6mm-FWHM Gaussian filtering on the reconstruction.

The acquisition-reconstruction procedure was repeated on both PETs after transferring the phantom solution into a container, pouring an additional 300ml DTPA water solution into the container, mixing the container solution, and pouring back the phantom with this new solution, resulting in a 50 Gy mean absorbed dose in the sectors. The transferring and pouring was performed using the same syringe system than for the first acquisition.

Afterwards, the phantom was again acquired after a 10 fold activity decay, i.e. 5 Gy mean absorbed dose in the sectors, but with a 10 fold longer acquisition time, i.e. 200min per bed position.

The natural crystal radioactivity generates 1400 randoms/sec in the whole field of view (FOV) [13], taking into account the ring diameter, this gives $\approx$ 30 randoms/sec/dm$^3$ everywhere in the FOV.

The acquisition rate for a $^{90}$Y vial is 800 trues/sec/GBq [13]. One dm$^3$ with a 1 Gy absorbed dose corresponds to 0.02 GBq (assuming a density of 1kg/dm$^3$), i.e. $\approx$ 16 trues/sec/dm$^3$. For or a mean attenuation length of 25cm, i.e. corresponding to a thin patient, this results in $\approx$1.7 trues/sec/dm$^3$.

 Therefore, the $^{90}$Y-trues count rate become lower than the  $^{176}$Lu-randoms one in regions having absorbed dose below 18Gy.

For comparison purpose, the hot rod sectors was also acquired in a clinical $^{18}$FDG whole-body (WB) setup, i.e. filled with a $^{18}$F-FDG solution corresponding to a mean liver SUV=2 in a 300MBq 75kg-patient WB study and acquired with 1.5 min per bed position.

Circular region of interest (ROI) with diameter equal to that of the rod was drawn on each rod position. Afterwards the mean and variance of the counts per pixel in the ROIs were computed and normalized to obtain 100 for the largest hot rod sector of each setup. This methods allows to get free of activity assessment incertitude that could impact the $^{18}$F-$^{90}$Y comparison.

*Dosimetry assessment*

The voxel dose histograms were computed using a validated scheme [12,13]. In summary: an expectation maximization (EM) based spatial resolution recovery was applied to the reconstructed slices. This spatial resolution correction was shown to provide similar recovery coefficients than those of other PET systems including a point spread function (PSF) modelling in the reconstruction [12,13,20]. Last, the activity distribution was convolved with the $^{90}$Y dose kernel in water taking into account the continuous beta energy spectrum. For sake of presentation clarity, all reconstructed images were rescaled to the first acquisition time (decay correction lower than 3%).

Sector TOF-PET based EUDs were computed in the same way as in ref (34), i.e. using the 550ps TOF-PET/CT with the HCC cells radiosensitivity set to $\alpha=0.035Gy^{-1}$ and using the Jones and Hoban formalism (38):

$$EUD = -\frac{1}{\alpha} \ln\left(\frac{\sum_i e^{-\alpha D_i}}{N}\right) \qquad (1)$$

where $D_i$ is the absorbed dose in the voxel i inside the considered sector ROI and $N$ is the number of voxels contained in this ROI. The ROIs were dimensioned in order to have for each sector a ratio of 0.227 between the total rods area and the ROI area.

For validation, the actual EUD was also computed for the six hot rod sectors: the binary sector map (1x1x1mm$^3$-voxel) was 3D convolved with the $^{90}$Y dose deposition kernel rescaled in order to get a mean dose of 50 Gy and of 100 Gy in the sectors. The α value in this EUD computation was fit in order to match the TOF-PET based EUDs.

**Results**

Figure 2 shows hot rods slices reconstructed from the different acquisitions. The 100Gy-setup (D,E) provided results almost as good as what is obtained in $^{18}$F phantoms (G,H): rods of 4 and 5 sectors are individually visualized using the 550 and 320ps TOF-PET, respectively. Acquisition time increase and slices summation in the 5Gy-setup in order to get the same total number of recorded coincidences resulted in poor quality images (C,F).

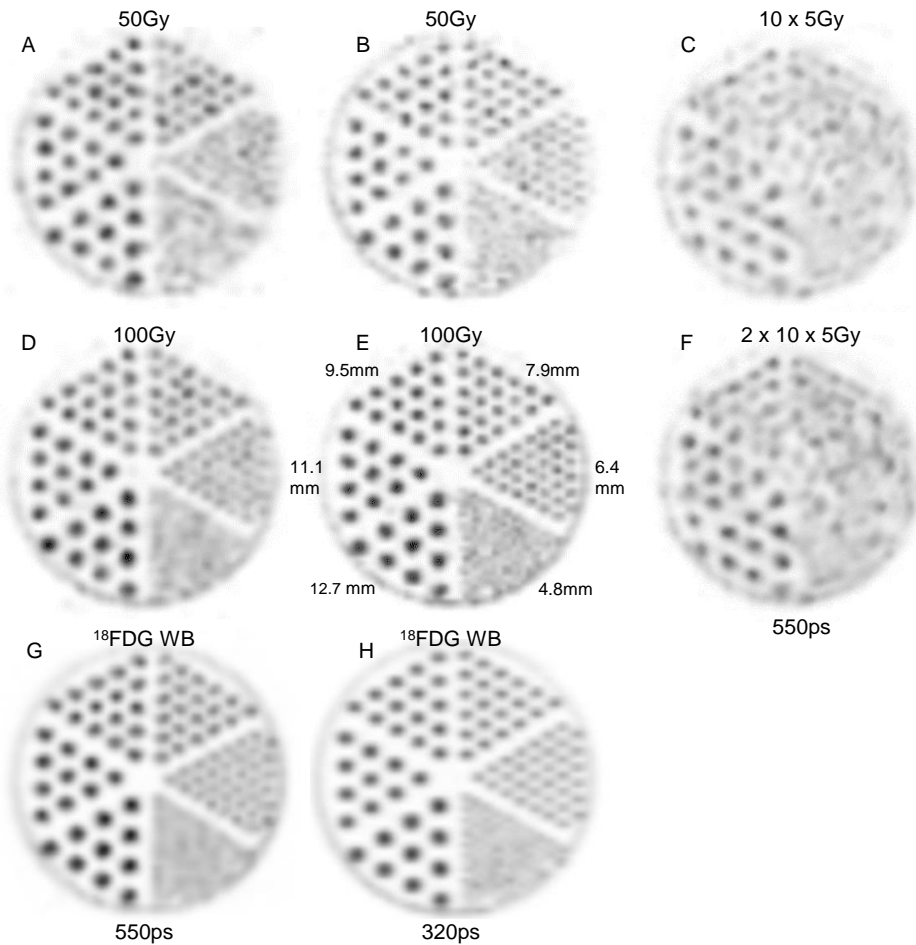

FIGURE 2: hot rod slice. A,C,D,F,G: 550ps TOF-PET. B,E,H: 320ps TOF-PET. A,B: 4mm-thick slice in 50Gy-$^{90}$Y-setup. D,E: 4mm-thick slice in 100Gy-$^{90}$Y-setup. C,F: 4mm-, 8mm-thick slice in 5Gy-$^{90}$Y-setup with a 10 fold longer acquisition time. Note that the acquisition time increasing and slice summation (C,F) do not counterbalance the count rate reduction and result in lower image quality than (A,D). G,H: 4mm-thick slice in clinical $^{18}$FDG WB setup.

Table 1 shows the normalized counts per pixel for the five largest hot rod sectors.

Table 1: mean and variance of the counts per pixel normalized to 100 for the largest hot rod sector of each setup for the two different TOF resolution (TOFr) systems..

| TOFr | Phantom | 12.7 mm | 11.1 mm | 9.5 mm | 7.9 mm | 6.4 mm |
|------|---------|---------|---------|--------|--------|--------|
| 550 ps | D: $^{90}$Y 100Gy | 100.0 ± 15.9 | 96.2 ± 10.9 | 80.8 ± 11.9 | 68.2 ± 13.0 | 52.2 ± 8.9 |
| | G: $^{18}$F DG WB | 100.0 ± 5.5 | 89.0 ± 5.4 | 77.9 ± 6.1 | 63.1 ± 5.1 | 48.7 ± 5.3 |
| 320 ps | D: $^{90}$Y 100Gy | 100.0 ± 7.7 | 109.5 ± 11.5 | 84.6 ± 7.7 | 68.7 ± 10.8 | 62.2 ± 11.6 |
| | G: $^{18}$F DG WB | 100.0 ± 7.1 | 89.5 ± 10.1 | 82.2 ± 10.5 | 68.0 ± 6.1 | 51.4 ± 5.4 |

Figure 3A shows the two devices dose-response reunification obtained using 550ps TOF-PET based EUD with a common 40Gy-dose threshold in line with what is observed in EBRT (34). For rods diameter above 9mm, figure 3B shows a good agreement between the 550ps TOF-PET based EUD (triangles) and the true sectors EUD computed with $\alpha$=0.021Gy$^{-1}$ (circles). This $\alpha$ value is thus the rescaling at the cell level of the apparent value 0.035Gy$^{-1}$ observed in 550ps TOF-PET (34) and is in line with the range observed in EBRT (see table 1). For smaller rod diameters a divergence is observed due to limited resolution recovery. The better agreement obtained for the smallest rod diameters likely resulted from the higher noise level observed in this sector (Fig. 2). The agreement quality is reduced by the Gaussian filtering (diamonds).

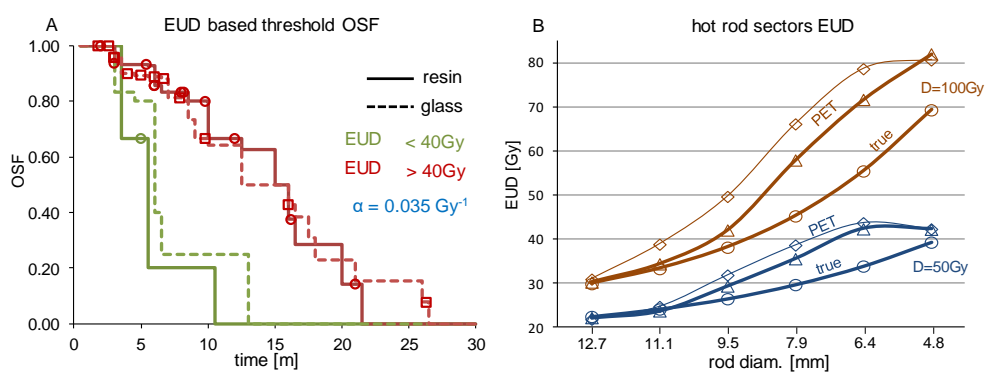

FIGURE 3: A: overall survival fraction (OSF) in HCC liver radioembolizations with resin spheres (solid line) and glass spheres (dashed line) using the same 40Gy-dose threshold on the 550ps TOF-PET based EUD:$\alpha$=0.035Gy$^{-1}$ (reprinted from (34) with permission of IOP publishing). B: comparison of the 550ps TOF-PET based sectors EUD:$\alpha$=0.035Gy$^{-1}$ (triangles, diamonds are without and with 6mm-FWHM filtering, respectively) with the true sectors EUD:$\alpha$=0.021Gy$^{-1}$(circles) for a mean sector dose D of 50Gy (blue) and of 100Gy (brown).

**Discussion**

To the best of our knowledge, it is the first time that a phantom study was conducted in order to evaluate whether valuable information from heterogeneous sources can be retrieved from $^{90}$Y PET imaging. We performed this analysis via the EUD formalism well adapted to predict radiobiological effects in $^{90}$Y radioembolization (34). Indeed, the low number of spheres results in macroscopic heterogeneity patterns (Fig. 1C). The impact of the heterogeneous activity distribution on the efficacy per Gy reported for high and low specific activity spheres was initially proved by Monte Carlo (41,42) and brilliantly confirmed in an animal model (43).

Figure 1C (35) clearly shows that the actual activity distribution post radioembolization is a mix of millimetric to centimetric heterogeneity pattern scales. Note that if the pattern scale (i.e. the inter-rods distance in the sectors, or equivalently the mean distance between sphere clusters in a tissue region) becomes lower than 4 mm, than the actual dose distribution tends to be uniform due to the $^{90}$Y beta range (4.3 mm mean range in water). This feature explains why the PET based EUD overestimation is maximal ($\approx$25%) around 6 mm (Fig. 3B) and afterwards reduces when the activity pattern scale decreases.

The present results clearly show that the EUD behaviour versus the heterogeneity scale is qualitatively reproduced using TOF-PET reconstruction with spatial resolution recovery and without any dedicated noise filtering. This observation explains why it was possible to reunify patient survival fraction of glass and resin sphere radioembolization using TOF-PET based EUD (34). The $\alpha$ value used in the actual EUD computation was reduced a little bit to fit the PET based EUD. This reduction results from the incomplete spatial resolution recovery joined with the reconstruction voxel size (4x4x4mm$^3$). Therefore, this reduced value $\alpha$=0.021Gy$^{-1}$ can be seen as the intrinsic HCC cell radiosensitivity obtained in vivo via TOF-PET based EUD and is in line with the intrinsic range reported in EBRT (see table 1), i.e. 0.01-0.04 Gy$^{-1}$ (39).

The phantom reconstructions (Fig.2) showed that $^{90}$Y TOF-PET in clinical count rates provides an image qualitatively almost as good as that obtained in $^{18}$F phantoms. Table 1 shows that rods variance of the $^{90}$Y 100Gy setup was about 2.5 fold that of the $^{18}$FDG WB setup using the 550ps TOF-PET, and almost similar using the 320ps TOF-PET, but for both systems the specific counts for the $^{90}$Y 100Gy setup had a slower decrease with the rods diameter than that of the $^{18}$FDG WB setup, supporting thus a little bit better spatial resolution for $^{90}$Y!  This amazing feature remains unexplained regarding that the positron energy of $^{90}$Y (mean=435 keV, max=739 keV (44)) is higher than that of $^{18}$F (mean=250 keV, max=634 keV).

On the other hand, due to the constant random rate generated by the natural LYSO radioactivity, a reduction of the specific activity cannot be counterbalanced by increasing the acquisition time and by summing slices (24-27).  Indeed, below 18Gy the imaging count rate mainly originates from the crystal radioactivity. Fortunately, efficient or toxic absorbed doses in radioembolization are above 40Gy, thus in a range where the crystal radioactivity has a low impact on TOF-PET imaging. Therefore, targets absorbed dose in phantoms modelling $^{90}$Y TOF-PET post liver radioembolization imaging should always be above 40Gy and clearly be reported by the corresponding delivered dose [Gy] to the volumes of interest.

Beside the present results showing that $^{90}$Y TOF-PET post liver radioembolization does not require any special filtering in tumour EUD assessment, independent methodologies already proved that the activity heterogeneity observed in $^{90}$Y TOF-PET imaging of the normal liver tissue also reflects the actual spheres distribution, and not noise artefact. Figure 4 gives an overview of these studies which are namely: Monte Carlo simulation of the spheres transport along the arterial hepatic tree (41,45), autoradiography of resected liver tissue post radioembolization (35) and in vivo MRI imaging post $^{166}$Ho liver radioembolization (46).

The present study suffers from the limitation that rod sectors were used rather than grid distribution of spheres. However this issue is mitigated for the 100Gy setup. Indeed only 18 mm of the rods length were filled, thus for the 12.7 mm diameter sector the distribution is closer to spheres than to rods. This issue could be fully solved for any diameters in further studies using 3D printed phantoms that could better models the activity heterogeneities observed in tumour and normal liver tissue.

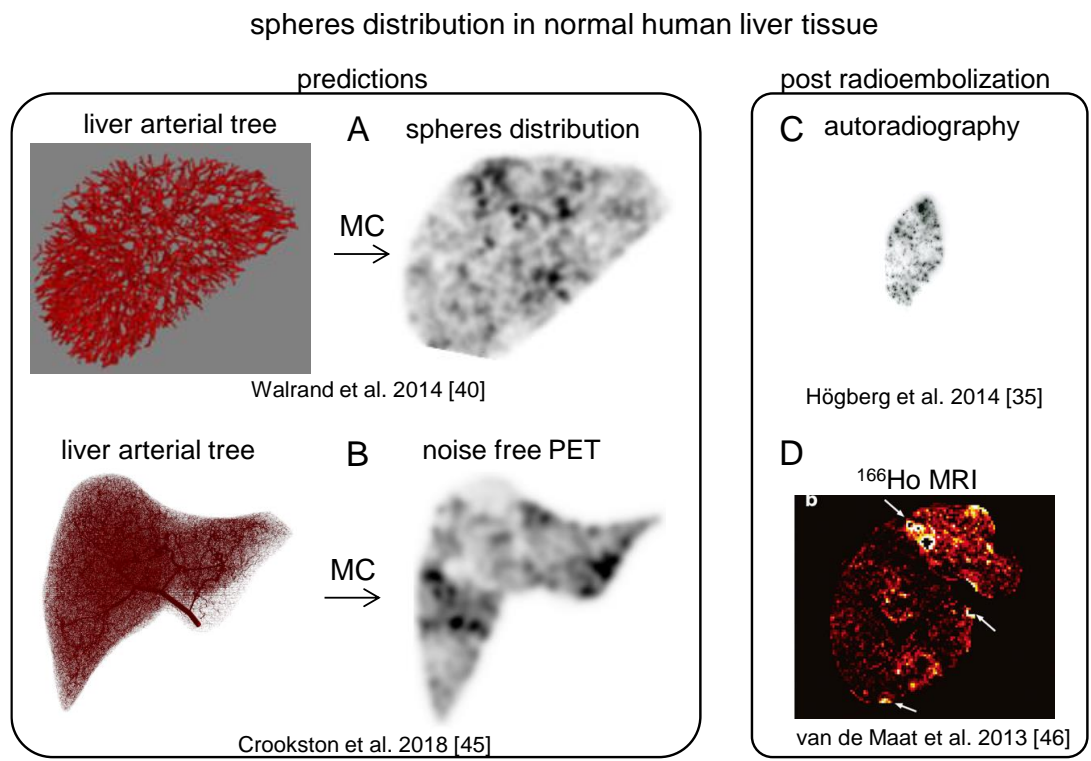

FIGURE 4: overview of studies validating the huge sphere distribution heterogeneity observed post liver radioembolization. Left panel: Monte Carlo simulations of the spheres transport inside synthetic hepatic arterial trees (reprinted from (41) with permission of the SNM and (45) with permission of IEEE). Right panel: autoradiography of a normal liver (NL) tissue resected 9 days post $^{90}$Y resin spheres liver radioembolization delivering 52Gy to the NL tissue (reprinted from (35) with permission of EJNMMI) and MRI imaging post $^{166}$Ho

loaded spheres liver radioembolization (reprinted from (46) with the permission of the authors). Images were represented with a similar scale.

## Conclusions

Standard FDG reconstruction parameters are suitable in TOF-PET post [90]Y liver radioembolization for accurate tumor EUD computation and normal liver tissue activity distribution assessment. EUD is the formalism of choice to take into account distribution heterogeneities due to variable microspheres decaying activities and to differences in tumor vascularization.

The present phantoms imaging clearly rules out the use of low specific activity phantom studies, i.e. corresponding to absorbed doses lower than 40Gy, aiming to optimize reconstruction parameters in TOF-PET imaging post [90]Y liver radioembolization. Indeed, increasing the acquisition time cannot counterbalance the noise resulting from the constant randoms rate originating from the natural crystal radioactivity.

## Compliance with Ethical Standards:

This study did not receive any funding

The authors declare that they have no conflict of interest related to the study

No animal and no human was involved in this study carcinoma patients treated with 90Y-loaded glass microsphere radioembolization. Liver International. 37:101-10.

**Figures caption:**

FIGURE 1: A) Jaszczak deluxe phantom set in vertical position on a paper bloc modelling the patient attenuation. Only a part of the hot rod insert was filled with active solution in order to reach a typical clinical tumour absorbed dose using a 3GBq total 90Y activity. B) Actual hot rods map. C) Autoradiography of a normal liver (NL) tissue resected 9 days post 90Y resin spheres liver radioembolization delivering 52Gy to the NL tissue (reprinted from (35) with permission of EJNMMI). B and C are represented at the same scale.

FIGURE 2: hot rod slice. A,C,D,F,G: 550ps TOF-PET. B,E,H: 320ps TOF-PET.  A,B: 4mm-thick slice in 50Gy-90Y-setup.  D,E: 4mm-thick slice in 100Gy-90Y-setup. C,F: 4mm-, 8mm-thick slice in 5Gy-90Y-setup with a 10 fold longer acquisition time. Note that the acquisition time increasing and slice summation

(C,F) do not counterbalance the count rate reduction and result in lower image quality than (A,D). G,H: 4mm-thick slice in clinical $^{18}$FDG WB setup.

FIGURE 3: A: overall survival fraction (OSF) in HCC liver radioembolizations with resin spheres (solid line) and glass spheres (dashed line) using the same 40Gy-dose threshold on the 550ps TOF-PET based EUD:α=0.035Gy$^{-1}$ (reprinted from (34) with permission of IOP publishing). B: comparison of the 550ps TOF-PET based sectors EUD:α=0.035Gy$^{-1}$ (triangles, diamonds are without and with 6mm-FWHM filtering, respectively) with the true sectors EUD:α=0.021Gy$^{-1}$(circles) for a mean sector dose D of 50Gy (blue) and of 100Gy (brown).

FIGURE 4: overview of studies validating the huge sphere distribution heterogeneity observed post liver radioembolization. Left panel: Monte Carlo simulations of the spheres transport inside synthetic hepatic arterial trees (reprinted from (41) with permission of the SNM and (45) with permission of IEEE). Right panel: autoradiography of a normal liver (NL) tissue resected 9 days post $^{90}$Y resin spheres liver radioembolization delivering 52Gy to the NL tissue (reprinted from (35) with permission of EJNMMI) and MRI imaging post $^{166}$Ho